\documentclass[twocolumn]{jpsj3}
\usepackage{txfonts}
\usepackage{epstopdf}

\title{Frequencies of the Edge-Magnetoplasmon Excitations in Gated Quantum Hall Edges}

\author{Akira Endo\thanks{akrendo@issp.u-tokyo.ac.jp}, Keita Koike, Shingo Katsumoto, and Yasuhiro Iye}
\inst{Institute for Solid State Physics, University of Tokyo, 5-1-5 Kashiwanoha, Kashiwa, Chiba 277-8581, Japan} 

\abst{We have investigated microwave transmission through the edge of quantum Hall systems by employing a coplanar waveguide (CPW) fabricated on the surface of a GaAs/AlGaAs two-dimensional electron gas (2DEG) wafer. An edge is introduced to the slot region of the CPW by applying a negative bias $V_\mathrm{g}$ to the central electrode (CE) and depleting the 2DEG below the CE\@. We observe peaks  attributable to the excitation of edge magnetoplasmons (EMP) at a fundamental frequency $f_0$ and at its harmonics $i f_0$ ($i=2$, 3,...). The frequency $f_0$ increases with decreasing $V_\mathrm{g}$, indicating that EMP propagates with higher velocity for more negative $V_\mathrm{g}$. The dependence of $f_0$ on $V_\mathrm{g}$ is interpreted in terms of the variation in the distance between the edge state and the CE, which alters the velocity by varying the capacitive coupling between them. The peaks are observed to continue, albeit with less clarity, up to the regions of $V_\mathrm{g}$ where 2DEG still remains below the CE.}

\begin{document}
\maketitle

\section{Introduction}
A coplanar waveguide (CPW) \cite{Wen69} placed on the surface of a GaAs/AlGaAs two-dimensional electron gas (2DEG) wafer has been employed to explore various aspects of the underlying 2DEG\@. The interaction of the microwave propagating through the CPW with the 2DEG provides us with a wealth of information on the properties of the 2DEG and on the high-frequency phenomena that take place in the 2DEG\@. 

The microwave is absorbed by the 2DEG located beneath the slots of the CPW\@. The intensity of the absorption increases with the increase of the 2DEG conductivity, allowing us to deduce the conductivity from the microwave transmission. The measurements of the transmission have thus been applied, for instance, to study the finite-frequency scaling of the conductivity in the integer quantum Hall (IQH) effect regime \cite{Engel93} and to observe the commensurability oscillations in the conductivity (as opposed to more usual observation in the resistivity) in unidirectional lateral superlattices \cite{Kajioka11, KajiokaCO13}. 

The absorption of the microwave leads, at the same time, to the local heating of the 2DEG residing below the slots. This enables us to introduce the temperature gradient into the 2DEG, with which we can also measure the thermoelectric voltages. Provided that the microwave power is not too high, the absorption raises only the electron temperature, leaving the lattice temperature intact. This allows us to selectively measure the thermoelectric voltages due to the diffusion contribution, eliminating the phonon-drag contribution. Note that, in a GaAs/AlGaAs 2DEG embedded in a wafer with the thickness of hundreds of microns, the latter contribution often dominates the thermoelectric voltages measured employing an external heater to introduce the temperature gradient \cite{Fletcher86}. The method has been applied to measure the diffusion thermoelectric voltages in the quantum Hall systems in the Corbino geometry.\cite{Kobayakawa13} 

The microwave can also excite collective modes in the quantum Hall systems. In fact, microwave transmission through the CPW has been extensively used as a tool to study pinning modes of varieties of electron-solid-like states in the quantum Hall systems: Wigner crystals both in the low filling of the lowest Landau level \cite{Ye02} and in the close vicinity of integer fillings,\cite{Chen03} and bubble \cite{Lewis02, Lewis04} and stripe \cite{Sambandamurthy08, Zhu09} phases in partially filled high Landau levels. 

In the present paper, we apply measurements of microwave transmission through CPW to investigate the excitations at the edges of a 2DEG in the quantum Hall states. Edge states in the quantum Hall systems have been a subject of long-standing interest since the time their crucial role in the quantum Hall effect was recognized \cite{Halperin82}. They are still attracting wide interest as dissipationless chiral channels capable of carrying charges or spins without backscattering (see, e.g., Refs. \citen{Stace04,Elman17,Hashisaka17}),  and accordingly as a prototype of chiral edge states in the extensively studied topological insulators \cite{Hasan10R}. An important collective excitation that takes place at the edges of a 2DEG is the edge magnetoplasmons (EMP) \cite{Fetter85,Volkov85,Volkov88}, a resonance in plasma oscillations having a much lower resonant frequency compared with the bulk counterpart. Extensive experimental studies have been done both in the frequency domain \cite{Allen83,Andrei88,Talyanskii90,Grodnensky90,Wassermeier90,Talyanskii92,Mahoney17} and in the time domain \cite{Ashoori92,Zhitenev95,Sukhodub04,Kamata10,Kumada11}. We target EMP in the present work.

In the present study, edges are electrostatically generated by applying a negative bias $V_\mathrm{g}$ to the central electrode (CE) of the CPW\@.  Edges are thus introduced into the slot region, where the CPW  measurements have high sensitivity. We observe, at IQH states ranging from the filling factor $\nu_0 = 2$ to 18, peaks at a series of frequencies $i f_0$ ($i = 1$, 2, 3,...) in the microwave transmission and also in the concomitantly measured thermoelectric voltages between Ohmic contacts. The peaks are attributable to the excitation of the fundamental mode ($i =1$) and the higher harmonics ($i = 2$, 3,...) of EMP\cite{Volkov88,Allen83,Andrei88,Talyanskii90,Grodnensky90,Wassermeier90,Talyanskii92}. Among different IQH states, $f_0$ takes a higher value for a higher $\nu_0$.  The peaks are observed both in the regime where the area under the CE is completely depleted ($V_\mathrm{g} < V_\mathrm{dpl}$) and in the regime where the 2DEG remains beneath CE with reduced density ($V_\mathrm{dpl} < V_\mathrm{g} < 0$), and the fundamental frequency $f_0$ is found to depend on $V_\mathrm{g}$ in both regimes. Within a single quantum Hall plateau, $f_0$ is found to increase with a decreasing magnetic field $B$. We will present semi-quantitative explanation for the dependence of $f_0$ on $V_g$ and on $B$ in $V_\mathrm{g} < V_\mathrm{dpl}$, and a qualitative account for the behavior in $V_\mathrm{dpl} < V_\mathrm{g} < 0$.

\section{Experimental Details \label{exp}} 
\begin{figure}[tb]
\includegraphics[width=8.5cm]{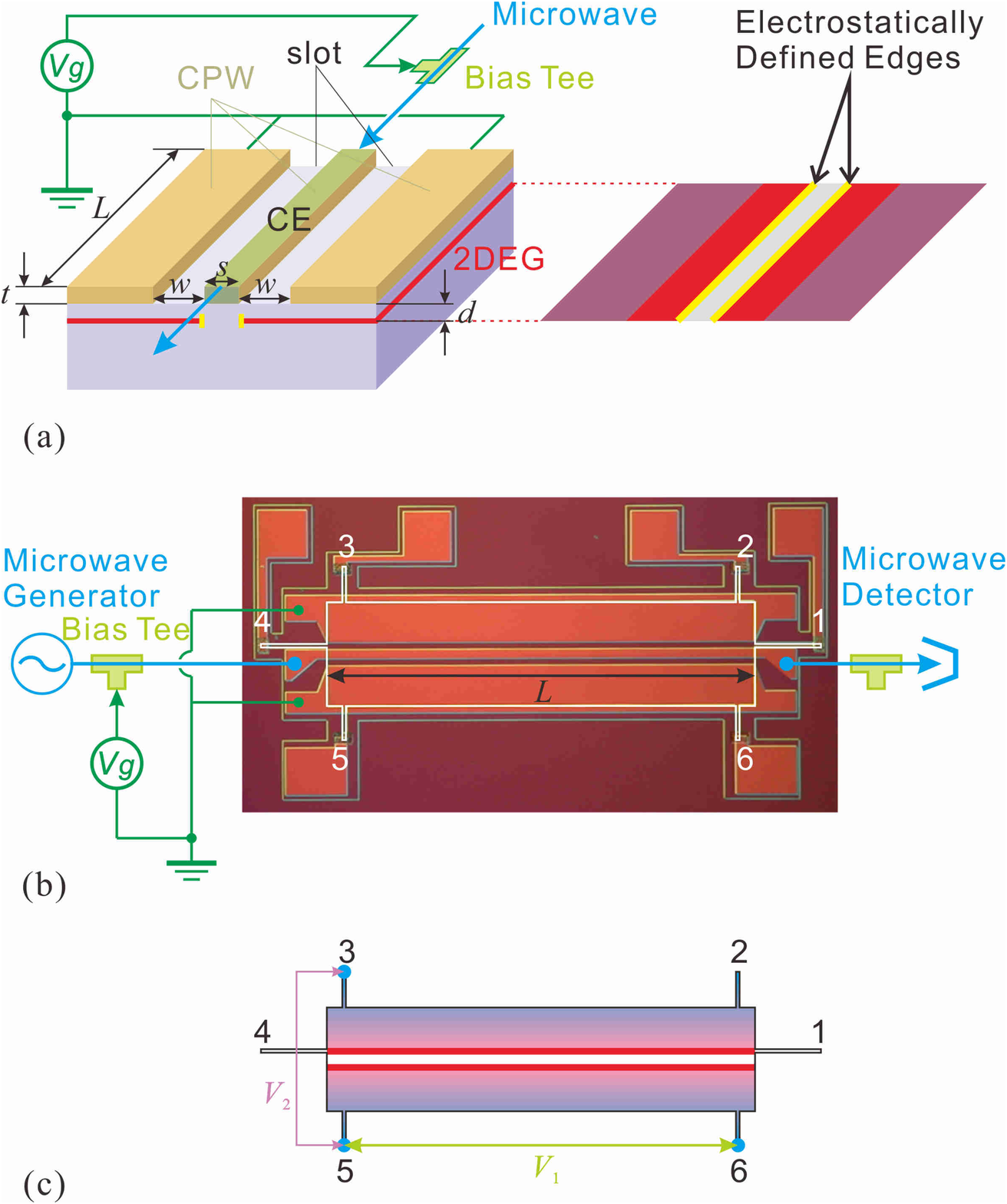}%
\caption{\label{samples} (Color online) (a) Schematic drawing of the sample used in the measurements. A coplanar waveguide (CPW) designed to have the characteristic impedance $Z_0 = 50$ $\Omega$ is placed on the surface of a 2DEG wafer. While measuring the transmission through the CPW,  electrostatically defined edge states can be generated near the edges of the slot regions by applying a negative bias $V_\mathrm{g}$ to the central electrode (CE) via a bias tee. The dimensions of the CPW are as follows: CE width  $s = 40$ $\mu$m, slot width $w = 28$ $\mu$m,  thickness $t = 60$ nm, and length $L =1 .8$ mm. The 2DEG is located at the depth $d = 65$ below the surface. (b) Optical micrograph of the device, with the schematics of the wiring. The 2DEG resides in the mesa area enclosed by white lines, and contains six Ohmic contacts labeled by numbers 1--6 in the figure. (c) Schematics of the measurement  of the thermoelectric voltages, $V_1$ and $V_2$. Microwaves are absorbed selectively at the slot regions and locally heats the electrons, thereby introducing the electron-temperature gradient.}
\end{figure}

The device used in the present study and the procedure of the measurement are schematically illustrated in Fig.\ \ref{samples}. A CPW designed to have the characteristic impedance $Z_0 = 50$ $\Omega$ is fabricated on the surface of a GaAs/AlGaAs 2DEG wafer with a Au/Ti film by using electron-beam lithography. Microwaves generated with a signal generator (Rhode \& Schwarz SMB100A) are injected into the CPW, and after propagating above the 2DEG, detected by a diode detector (Keysight 8474E) (see Figs.\ \ref{samples}(a) and \ref{samples}(b)). The data presented in this paper were taken using the source microwave power of $-40$ dBm.
The mobility and the electron density of the 2DEG used in the device was $\mu = 80$ m$^2$V$^{-1}$s$^{-1}$ and $n_{0} = 4.32 \times 10^{15}$ m$^{-2}$, respectively, and the 2DEG is located at the depth $d = 65$ nm from the surface \cite{2DEGdepth}. 
The wafer is mesa-etched and the 2DEG resides within the area delineated by the white lines in Fig.\ \ref{samples}(b). At the ends of the protruded arms, six Ohmic contacts are placed (labeled by the numbers 1--6 in Figs.\ \ref{samples}(b) and \ref{samples}(c)). As mentioned earlier, microwave absorption at the slot regions locally heats the electrons in these areas, resulting in the electron-temperature gradient towards the Ohmic contacts, as schematically depicted in Fig.\ \ref{samples}(c). Although the temperature gradient can have complicated spatial distribution, especially when placed in a strong magnetic field,\cite{Endo17SD} and the details are not well known, we can still measure thermoelectric voltages between Ohmic contacts, as exemplified by $V_1$ and $V_2$ in Fig.\ \ref{samples}(c). These are probably some mixture of the longitudinal (Seebeck, $S_{xx}$) and the transverse (Nernst, $S_{xy}$) components, with $V_1$ ($V_2$) mainly composed of $S_{xy}$ ($S_{xx}$). As we will see, this uncertainty does not affect the observation of the peaks. 
For both the microwave transmission and the thermoelectric voltage measurements, we used lock-in technique with low-frequency (17 Hz) amplitude modulation of the microwave source.

In order to introduce edges to the slot regions, a negative bias $V_\mathrm{g}$ is applied to the CE of the CPW via a bias-tee coupling (see Figs.\ \ref{samples}(a) and \ref{samples}(b)). The measurement was performed in a dilution refrigerator (Oxford TLD) equipped with a superconducting magnet and a pair of semi-rigid coaxial cables. The sample is immersed in the mixing chamber of the refrigerator held at the temperature of $\sim$20 mK\@.

\section{Results \label{Results}}
\subsection{Frequency and gate-voltage dependence at an integer quantum Hall state \label{secfVdep}}
\begin{figure*}[tb]
\includegraphics[width=19.0cm]{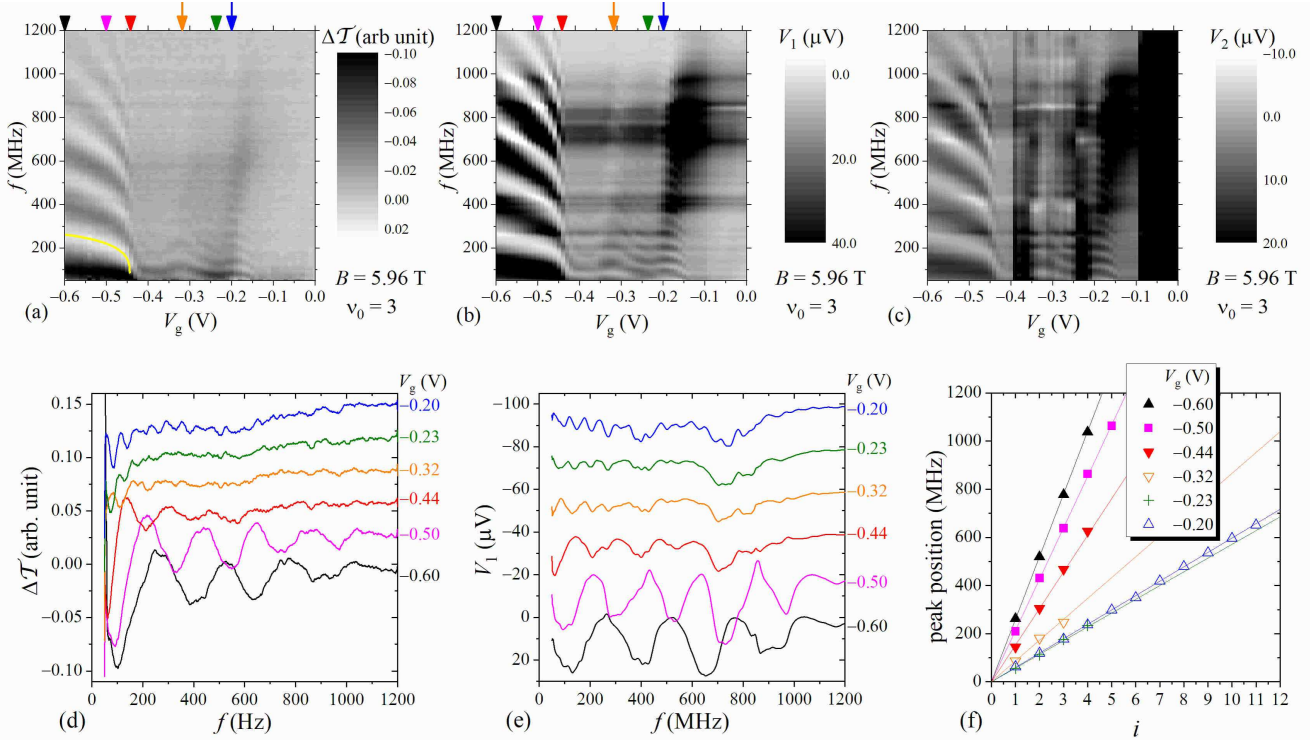}%
\caption{\label{nu3Vdep} (Color online) Measurement results at $\nu_0 =3$ IQH state ($B = 5.96$ T). (a) Transmission $\Delta \mathcal{T}(f;V_\mathrm{g}) = \mathcal{T}(f;V_\mathrm{g}) - \mathcal{T}(f;0)$. Values at $V_\mathrm{g} = 0$ is subtracted as a background. (b) Thermoelectric voltage between contacts 5 and 6, $ V_1(f;V_\mathrm{g})$, and (c) between contacts 5 and 3, $V_2(f;V_\mathrm{g}) $. (d) (e) Cross-sections of (a) and (b), respectively, at several fixed gate voltages $V_\mathrm{g}$ indicated by arrows or arrow heads having the same color. Traces are offset for clarity. (f) Positions of the $i$-th peak. Fundamental frequency $f_0$ can be obtained from the slope of the linear fittings shown by the lines in the figure. Thick yellow curve in (a) shows calculated $f_0(V_\mathrm{g})$ (see Sect.\ \ref{Calcf0}).}
\end{figure*}

First we describe typical behaviors we observe at an IQH state, taking the measurement performed at the filling factor $\nu_0 = n_0 h / (eB) = 3$ for example. Figure \ref{nu3Vdep}(a) shows a grayscale plot of the dependence of the microwave transmission through the CPW on the voltage $V_\mathrm{g}$ applied to the CE and on the frequency $f$ of the microwave. The data were taken by sweeping the frequency for a fixed $V_\mathrm{g}$, and then by varying the $V_\mathrm{g}$ stepwise from 0 V to $-$0.6 V with the interval of 0.01 V\@. The frequency sweeps for several selected values of $V_\mathrm{g}$ (indicated by arrows or arrow heads on the top of Fig.\ \ref{nu3Vdep}(a)) are plotted in Fig.\ \ref{nu3Vdep}(d). Since the measured transmission $\mathcal{T}(f;V_\mathrm{g})$ contains large background resulting from the frequency dependence of the transmission outside the sample (e.g., at the connection between the cable and the sample), transmission at $V_\mathrm{g} = 0$ V, $\mathcal{T}(f;0)$, was subtracted as the background in the plots to pick out the changes $\Delta \mathcal{T}(f;V_\mathrm{g})$ brought about by applying the negative $V_\mathrm{g}$. The density $n_\mathrm{CE}$ of the 2DEG under the CE decreases with decreasing $V_g$ and vanishes at $V_\mathrm{dpl} = -0.44$ V\@. In the region $V_\mathrm{g} < V_\mathrm{dpl}$, where the area below the CE is depleted, we can see a series of clear peaks, namely the enhancement of the transmission, in the plot of $\Delta \mathcal{T}(f;V_\mathrm{g})$. The peaks shift to higher $f$ side for more negative $V_\mathrm{g}$. Closer look of Fig.\ \ref{nu3Vdep}(a) reveals that the peaks continue, albeit with less clarity, up to the region $V_\mathrm{dpl} < V_\mathrm{g}$ with smaller frequencies. The peak slightly shifts to higher $f$ side at around $V_\mathrm{g} = -0.32$ V and $-0.20$ V, seen as humps in the figure (pointed by arrows on the top).

Essentially the same information can be obtained with the thermoelectric voltages between the Ohmic contacts, as can be seen in Figs.\ \ref{nu3Vdep}(b) and \ref{nu3Vdep}(c).  In these figures, the voltages $V_1$ and $V_2$ depicted in Fig.\ \ref{samples}(c) are plotted as a function of $f$ and $V_\mathrm{g}$. The sign of $V_1$  ($V_2$) is defined to be positive when the voltage at the contact 5 is higher than that at the contact 6 (3). 
The peaks are observed in Fig.\ \ref{nu3Vdep}(b) even clearer than in Fig.\ \ref{nu3Vdep}(a). Selected frequency sweeps for $V_1$ are presented in Fig.\ \ref{nu3Vdep}(e), showing the peaks at the same frequencies as in Fig.\ \ref{nu3Vdep}(d). An advantage of the thermoelectric voltage over $\mathcal{T}$ is that signals attributable to the sample alone can be obtained without the background subtraction. In this magnetic field ($\nu_0 = 3$), higher $\mathcal{T}$ corresponded to smaller $V_1$. Therefore, smaller $V_1$ is plotted with a lighter tone in Fig.\ \ref{nu3Vdep}(b) and upward in Fig.\ \ref{nu3Vdep}(e), in order to facilitate the comparison with Figs.\ \ref{nu3Vdep}(a) and \ref{nu3Vdep}(d), respectively. As we will see below, we found that the correspondence between the increase or decrease of $\mathcal{T}$ and that of $V_1$ depends on the magnetic field. 
The peaks are less apparent in Fig.\ \ref{nu3Vdep}(c). (Again, higher $\mathcal{T}$ corresponded to smaller $V_2$). Note that while $V_1$ is the voltage between the contacts residing on the same side (lower side in Fig.\ \ref{samples}(c)) of the CPW, $V_2$ straddles the CPW\@. Signals are noisy and less clear in $V_2$, probably because the temperature gradient is mostly compensated between the two sides of the CPW, and the voltage from the depleted area below the CE is included. We will not discuss the voltage $V_2$ any further.

The frequency of the $i$-th peak is plotted in Fig.\ \ref{nu3Vdep}(f). Different symbols correspond to different values of $V_\mathrm{g}$. For each $V_\mathrm{g}$, the peak frequencies fall on a line, namely, the peak frequency is given by the relation $f = i f_0$, and thus attributable to the fundamental mode ($i =1$) and higher harmonics ($i = 2$, 3...) of the EMP excitation. The mechanism through which the presence of EMP enhances $\mathcal{T}$ is currently not known. The fundamental frequency $f_0$ is the frequency of the lowest ($i = 1$) peak, and can also be derived from the slope of the linear fitting. We can track the variation of $f_0$ with $V_\mathrm{g}$ by following the lowest peak in Figs.\ \ref{nu3Vdep}(a)--\ref{nu3Vdep}(c).

\subsection{Comparison among different integer quantum Hall states \label{cmpQH}}
\begin{figure*}[tb]
\includegraphics[width=18.5cm]{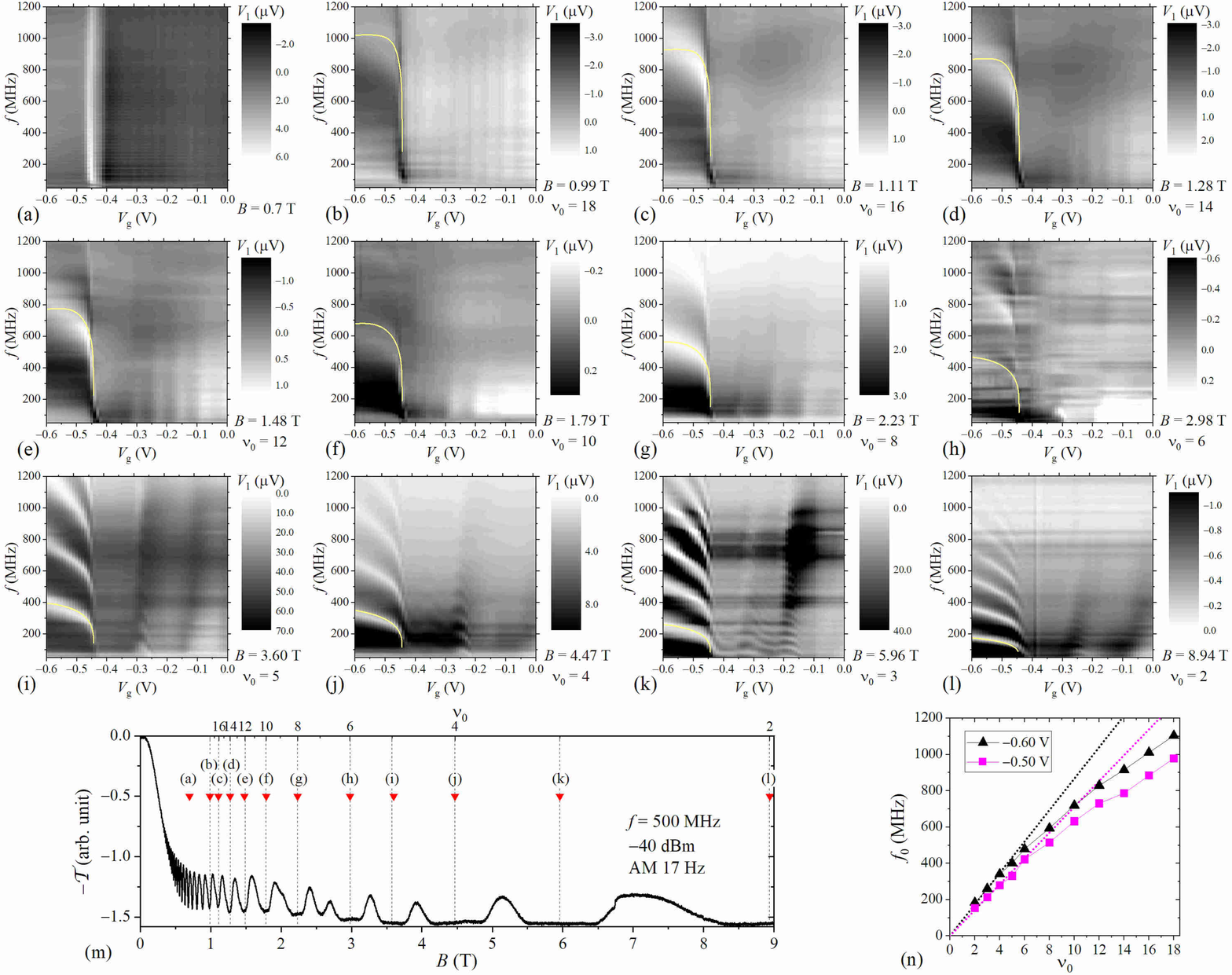}%
\caption{\label{IQHsVdep} (Color online) (a)--(l) $V_1(f;V_\mathrm{g})$ taken at $B = 0.7$ T (maximum of the Shubnikov-de Haas oscillations) (a) and at IQH states ranging from $\nu_0 =18$ to 2 (b)--(l). Yellow curves in (b)--(l) represent calculated $f_0(V_\mathrm{g})$ (see Sect.\ \ref{Calcf0}). (m) Magnetic-field dependence of $-\mathcal{T}(500 \mathrm{MHz};0 \mathrm{V})$, which roughly represents the lineshape of $\sigma_{xx}(B)$. The filling factor $\nu_0$ is presented in the top axis. Magnetic fields at which (a)--(l) were taken are indicated by downward arrowheads. (n) The fundamental frequency $f_0$ vs. the integer fillings $\nu_0$ for two different gate voltages $V_\mathrm{g}$. Dotted lines highlight the relation $f_0 \propto \nu_0$ for smaller $\nu_0$.}
\end{figure*}

Next we compare different IQH states. In Figs.\ \ref{IQHsVdep}(b)--\ref{IQHsVdep}(l), we display $V_1(f;V_\mathrm{g})$ taken at integer fillings ranging from $\nu_0 = 18$ to 2. They all show peak(s) for $V_\mathrm{g} < V_\mathrm{dpl}$ at progressively higher frequencies for higher fillings (lower magnetic fields). Peaks are not seen in Fig.\ \ref{IQHsVdep}(a) showing $V_1(f;V_\mathrm{g})$ collected at $B = 0.7$ T, where the quantum Hall state is not developed. (See Fig.\ \ref{IQHsVdep}(m) for the magnetic-field positions for (a)--(l).) The continuation of the peaks to $V_\mathrm{dpl} < V_\mathrm{g}$ are observed for fillings below $\nu_0 = 5$ accompanied by several humps (the shift of the peaks to higher $f$ side).  For higher fillings, vertical lines, increasing in number for increasing fillings, are discernible, which probably correspond to unresolved remnants of the humps. In Fig.\ \ref{IQHsVdep}(n), we plot the fundamental frequency $f_0$ against the filling factor $\nu_0$ for two different values of $V_\mathrm{g}$ in the depleted region. We can see that the relation $f_0 \propto \nu_0$ roughly holds for smaller $\nu_0$, but deviation to smaller $f_0$ becomes apparent for higher $\nu_0$.

As mentioned earlier, $V_1$ reflects $\mathcal{T}$. This can readily be understood considering that the change in $\mathcal{T}$ is caused by the change in the microwave absorption by the 2DEG in the slot region, which, at the same time, alters the temperature gradient introduced into the 2DEG mesa. However, the relation between $V_1$ and $\mathcal{T}$ is not straightforward. We found that the $V_1$ either increase or decrease with increasing $\mathcal{T}$ with differing conversion rate depending on the magnetic field. Owing to the intricate sample geometry, the temperature gradient has complicated spatial distribution, especially under a strong magnetic field \cite{Endo17SD}. Moreover, $S_{xy}$ can be either positive or negative depending on the magnetic field in the quantum Hall regime \cite{Jonson84}. Further complication can arise from possible reduction in the electron density beneath the grounded side gates of the CPW compared to density in the slot region owing to the contact potential difference \cite{contactpot}. Here and in what follows, $V_1$ is plotted with the choice of the sign that the higher $\mathcal{T}$ corresponds to a lighter tone (see the legends to the right of each figure), determined by comparing to the simultaneously measured $\mathcal{T}$. We have confirmed that  $V_1(f;V_\mathrm{g})$ plotted here basically reproduces $\Delta \mathcal{T}(f;V_\mathrm{g})$, but resolves the peaks more clearly especially in the region $V_\mathrm{dpl} < V_\mathrm{g}$ \cite{Suppl}. 

\subsection{Dependence on the magnetic field within a quantum Hall plateau \label{BdepQHP}}
\begin{figure}[tb]
\includegraphics[width=7.5cm]{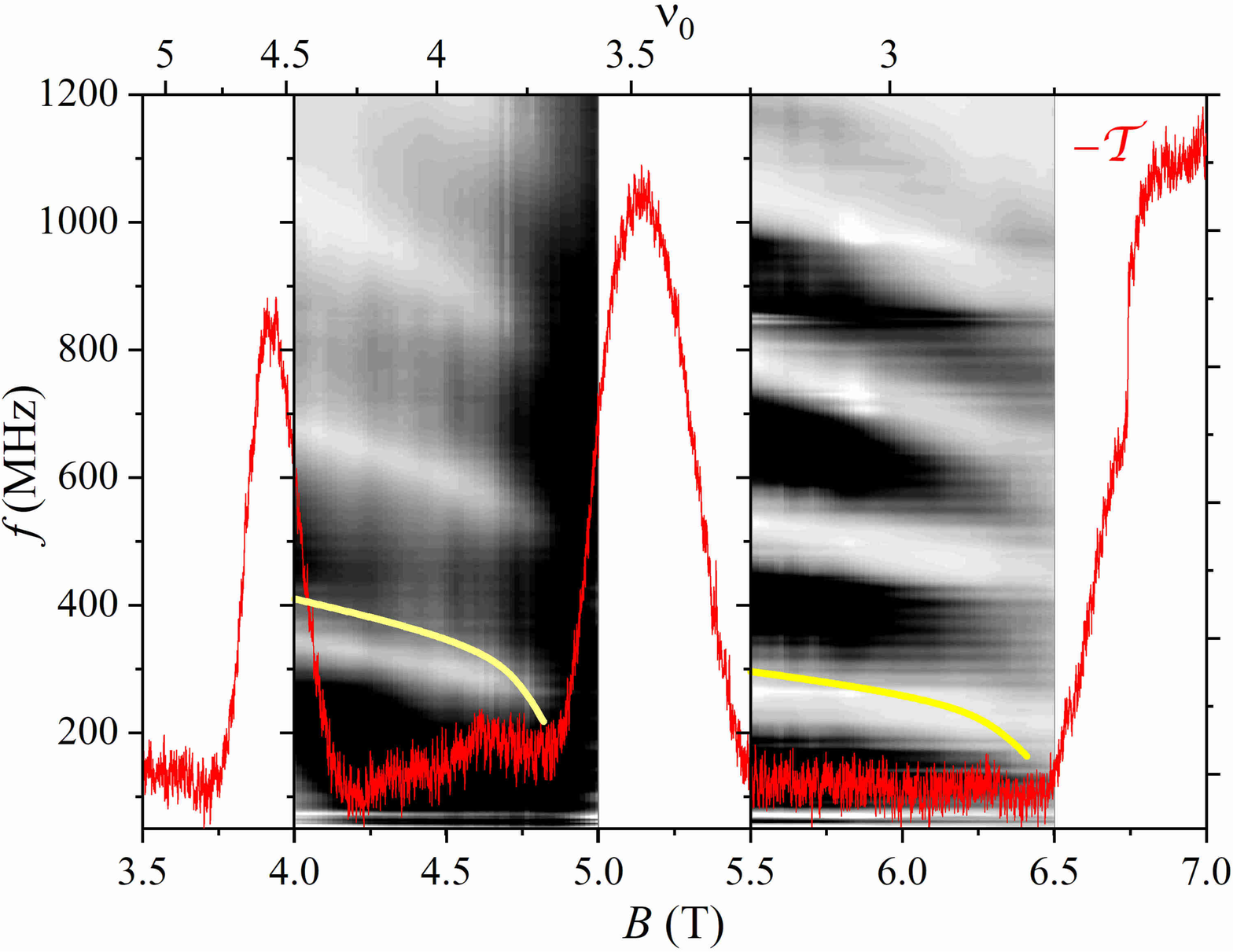}%
\caption{\label{Vm60nu34Bdep} (Color online) $V_1(f; -0.6\mathrm{ V})$ plotted in the $f$-$B$ plane for the magnetic-field ranges encompassing the IQH plateau regions around $\nu_0 = 4$ and 3. The filling factor $\nu_0$ is presented in the top axis. The magnetic-field dependence of $-\mathcal{T}(500 \mathrm{MHz};0 \mathrm{V})$, which basically represents $\sigma_{xx}(B)$, is also plotted with a thin (red) line. Thick yellow curves are calculated $B$-dependence of $f_0(-0.6 \mathrm{V})$ (see Sect.\ \ref{Calcf0}).}
\end{figure}
So far, we have examined the behaviors of $\Delta \mathcal{T}(f;V_\mathrm{g})$ and $V_1(f;V_\mathrm{g})$ at exact integer fillings. In this subsection, we see how they vary with the magnetic field within a quantum Hall plateau. Figure \ref{Vm60nu34Bdep} shows $V_1$ plotted in the $f$-$B$ plane for the magnetic fields encompassing the quantum Hall plateau region around the filling factors $\nu_0 = 3$ and 4. The CE is kept at $V_\mathrm{g} = -0.6$ V\@. The ranges of the plateaus are indicated by the simultaneously plotted $B$-dependence of $-\mathcal{T}$, which basically reproduces the lineshape of the longitudinal conductivity $\sigma_{xx}(B)$.  The positions of the peaks we see here, and hence the peaks in $\Delta \mathcal{T}$, slightly shifts to higher frequency side with the increase of $\nu_0$ (the decrease of $B$). 

\section{Discussion \label{Discussion}}
\subsection{Calculation of $f_0$ for $V_\mathrm{g} < V_\mathrm{dpl}$ \label{Calcf0}}
\begin{figure}[tb]
\begin{center}
\includegraphics[width=6.5cm]{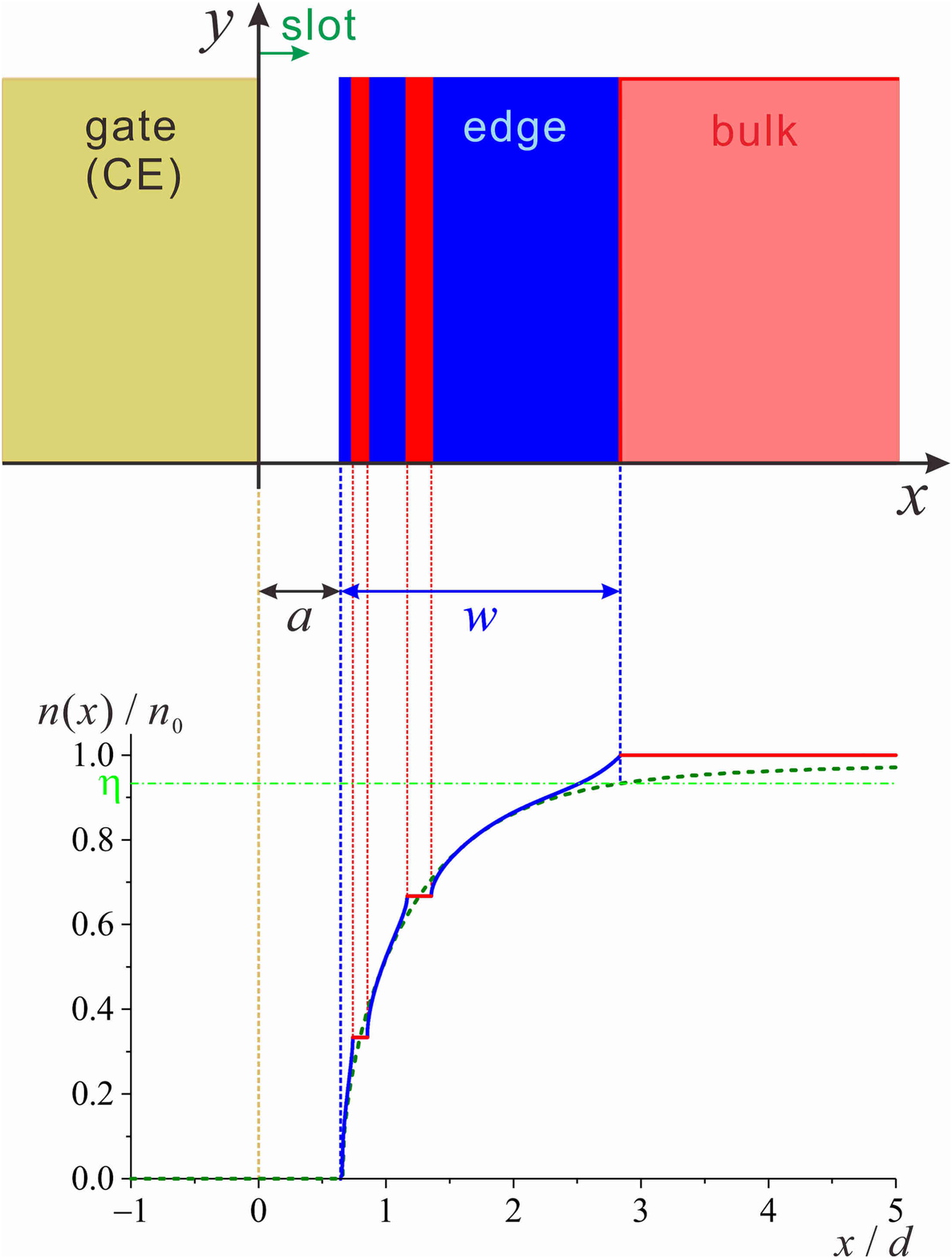}%
\caption{\label{nxdep} (Color online) Top: Topview of the gate ($x < 0$), depletion region ($0 < x < a$), edge region with compressible (blue) and incompressible (red) stripes ($a < x < a + w$), and bulk region ($a + w < x$). Bottom: Profile of the electron density $n(x)$ near the edge. Dashed and solid lines are for $B = 0$ T and the quantum Hall plateau region ($\nu_0 =3$), respectively. $\eta$ represents the separatrix for $n(x) / n_0$ dividing edge and bulk regions (see Eq.\ (\ref{wdef})).}
\end{center}
\end{figure}
\begin{figure}[tb]
\begin{center}
\includegraphics[width=7.5cm]{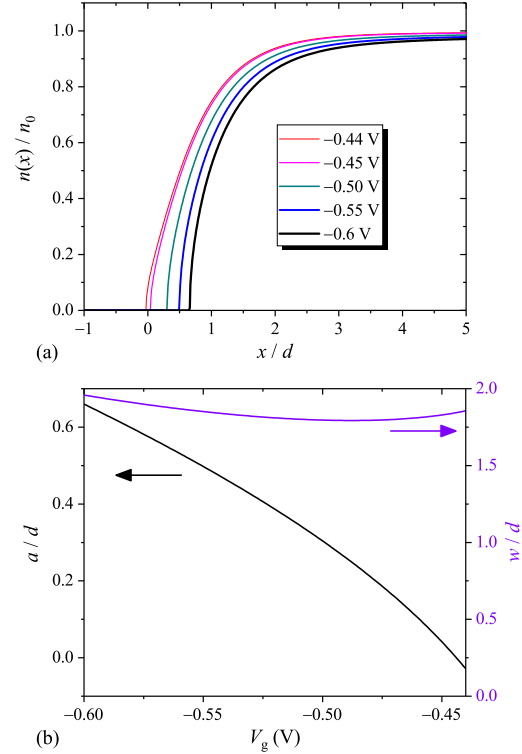}%
\caption{\label{nawVgdep} (Color online) (a) Profile of the electron density $n(x)$ near the edge for $V_\mathrm{g} = -0.44$, $-0.45$, $-0.50$, $-0.55$ and $-0.60$ V at $B = 0$ T\@. $n_0 = 4.32 \times 10^{15}$ m$^{-2}$ is the bulk electron density. (b) Depletion width $a$ (left axis), and the width of the edge region $w$ calculated with Eq.\ (\ref{wdef}) for $\nu_0 = p = 3$ (right axis). $d = 65$ nm is the depth of 2DEG from the surface.}
\end{center}
\end{figure}

In this section, we discuss the variation of the fundamental frequency, $f_0$, of the EMP excitation with the gate voltage $V_\mathrm{g}$. We first consider the range of the gate bias $V_\mathrm{g} < V_\mathrm{dpl}$, where the area beneath the CE is completely depleted. We calculate the dependence of $f_0$ on the gate bias $V_\mathrm{g}$, and then on the magnetic field $B$. To calculate $f_0$, we employ the local capacitance approximation \cite{Zhitenev95,Johnson03}, slightly modified for the present setup. In the original capacitance approximation \cite{Zhitenev95,Johnson03}, EMP of the 2DEG located below and facing a metallic gate is considered, while in the present case, the gate and the 2DEG is laterally separated by the depletion width $a$ (see Fig.\ \ref{nxdep}).
Following the same procedures as those used in the ordinary capacitance approximation, we combine the equation of continuity and the equation representing the capacitive coupling between the gate and the edge state, and obtain, 
\begin{equation}
f_0 = \frac{1}{L} \frac{\sigma_{xy}}{C} \label{f0CapMod}, 
\end{equation}
where $\sigma_{xy}$ is the Hall conductivity, $C$ the capacitance per unit length, and $L$ the length of the CPW (see Fig.\ \ref{samples}(b)). Derivation of Eq.\ (\ref{f0CapMod}) is given in Appendix \ref{Af0}. To make the calculation of $C$ analytically practicable, we assume that the gate and the edge state are two parallel stripes with the width $w$, located on the same plane side by side with the separation $a$ and neglect the depth $d$ of the 2DEG from the surface as was done in Ref.\ \citen{Chklovskii92}. (Note, however, that we neglected $d$ only in the calculation of $C$). Then we have \cite{Smythe50},
\begin{equation}
C = \epsilon \epsilon_0 \frac{K(\sqrt{1-k^2})}{K(k)}, \label{Capacitance}
\end{equation}
where $K(k)$ represents a complete elliptic integral of the first kind with $k = a /(a + 2 w)$, $\epsilon$ is the relative permittivity of the host crystal (we used the low-temperature value for GaAs $\epsilon =12.35$ given in the literature \cite{Strzalkowski76}), and $\epsilon_0 = 8.85 \times 10^{-12}$ F$\cdot$m$^{-1}$ is the vacuum permittivity.

Next, we calculate the depletion width $a$ and the width of the edge state $w$, using a theory by Larkin and Davies \cite{Larkin95}. We focus on the vicinity of the boundary between the gate (CE) and one of the adjacent slot areas where the edge states are formed. As depicted in Fig.\ \ref{nxdep}, we set the $x$-axis across the CPW and define the boundary as $x = 0$. The depletion width $a$ is the distance between $x = 0$ and the onset of the electron-density profile $n(x)$, and thus $n(a) = 0$. The theory \cite{Larkin95} provides analytical formulas for how $n(x)$ (without the magnetic field) and $a$ vary with $V_\mathrm{g}$ for a 2DEG residing at the depth $d$ from the surface. The formulas are given for two different models of the boundary condition at the surface of the 2DEG wafer, the ``pinned'' and ``frozen'' surface model. We employ the latter, which is expected to be a more realistic model at cryogenic temperatures \cite{pinned}. In Figs. \ref{nawVgdep}(a) and \ref{nawVgdep}(b), we plot $n(x)$ for several different values of $V_\mathrm{g} (\leq V_\mathrm{dpl})$ and $a$, respectively, calculated with these formulas using our sample parameters \cite{contactpot}. When the 2DEG is in the quantum Hall plateau region, the Fermi energy is located at the localized state near the tail of a disorder-broadened Landau level, and the filling factor of the bulk area $\nu_0 = n_0 h / (eB)$ resides within a narrow range encompassing an integer value, $p - \delta_- < \nu_0 < p+\delta_+$,  ($p = 1$, 2, 3,...). The filling factor near the edge, $\nu(x) = n(x) h / (eB)$, increases with $x$ and and enters the bulk localized area, $p - \delta_- < \nu(x)$, for large enough $x$. We define the edge-state width $w$ as the span of $x$ in which $\nu(x)$ varies from 0 to $p - \delta_-$. Thus we can obtain $w$ by solving
\begin{equation}
\frac{n(a+w)}{n_0} = \frac{\nu(a+w)}{\nu_0} = \frac{p-\delta_-}{\nu_0} \equiv \eta. \label{wdef}
\end{equation}
The width $w$ calculated using Eq.\ (\ref{wdef}) with $\nu_0 = p =3$ is also plotted in Fig.\ \ref{nawVgdep}(b) as a function of $V_\mathrm{g}$, where we used $\delta_- = 0.24$ deduced from the extent of the $\nu_0 = 3$ plateau seen in Fig.\ \ref{Vm60nu34Bdep} \cite{deltanote}. It is well known that in the quantum Hall regime, edge states are composed of alternating compressible and incompressible strips, and $n(x)$ is slightly altered from that for $B =0$ T through the redistribution of electrons around the incompressible strips as depicted in Fig.\ \ref{nxdep} \cite{Chklovskii92}. The width $w$ defined here represents the total width of the edge region containing both compressible and incompressible strips \cite{redisbulk}.

In Fig.\ \ref{nu3Vdep}(a), $f_0(V_\mathrm{g})$ at $\nu_0 = 3$ calculated by substituting the $a(V_\mathrm{g})$ and $w(V_\mathrm{g})$ shown in Fig.\ \ref{nawVgdep}(b) into Eq.\ (\ref{Capacitance}), and then the resulting $C$ into Eq.\ (\ref{f0CapMod}) along with $\sigma_{xy} = 3 e^2/h$, is plotted with the thick yellow line.  We can see that the calculated $f_0(V_\mathrm{g})$ reproduces the experimentally observed variation of $f_0$ with $V_\mathrm{g}$ well, except for the close vicinity of $V_\mathrm{g} \simeq V_\mathrm{dpl}$. The accuracy of our approximation neglecting the depth $d$ of the 2DEG in the calculation of $C$ deteriorates close to $V_\mathrm{dpl}$, where $a$ becomes small. 

\begin{figure}[tb]
\begin{center}
\includegraphics[width=6.5cm]{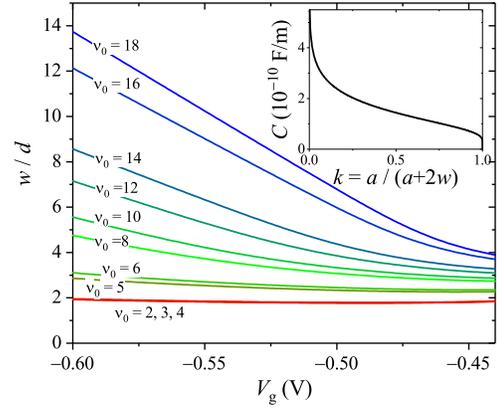}%
\caption{\label{wnus} (Color online) The width of the edge region $w$ as a function of $V_\mathrm{g}$ calculated by Eq.\ (\ref{wdef}) for various bulk integer filling factors $\nu_0$. Inset: the capacitance $C$ given by Eq.\ (\ref{Capacitance}) plotted against $k = a / (a+2w)$.}
\end{center}
\end{figure}

One can consider two different routes through which decreasing $V_\mathrm{g}$ leads, in principle, to the increase in $f_0$, or equivalently to the increase in the propagation velocity of the EMP \cite{Kamata10}.  First, the distance $a$ of the edge state from the gate (CE) increases with decreasing $V_\mathrm{g}$, which, in turn, increases the frequency (velocity) by diminishing the capacitive coupling to, and hence the screening by, the metallic gate. Second, decreasing $V_\mathrm{g}$ makes the confining potential at the edge steeper, and the resulting increase in the transverse electric field will also enhance the drift velocity along the edge in the magnetic field. Figure \ref{nawVgdep}(b) shows the expected increase in $a$ with decreasing $V_\mathrm{g}$. The steepening of the confining potential is reflected in the steeper rise of $n(x)$ around the onset of the population of the electrons for more negative $V_\mathrm{g}$ seen in Fig.\ \ref{nawVgdep}(a). For our EMP, however, we should consider the steepness over the width of the edge state. The steepness enhances $f_0$ through the narrowing of $w$, which reduces the capacitance $C$ (see Eqs.\ (\ref{f0CapMod}) and (\ref{Capacitance}), and the inset of Fig.\ \ref{wnus}). However, Fig.\ \ref{nawVgdep}(b) reveals that after initially exhibiting slight decrease on decreasing $V_\mathrm{g}$ from $V_\mathrm{dpl}$,  $w$ then shows mild upturn. Therefore, the main role of more negative $V_\mathrm{g}$ in the enhancement of $f_0$ is to repel the edge state farther away from the gate. 

The increase of $f_0$ with increasing integer fillings $\nu_0 =p$ ($p = 2$, 3, ... 18) demonstrated in Fig.\ \ref{IQHsVdep} are basically attributable to the relation $f_0 \propto \sigma_{yx} = p e^2/h$ in Eq.\ (\ref{f0CapMod}). The relation results in $f_0 \propto \nu_0 = p$ if $C$ does not vary with $p$. In Fig.\ \ref{wnus}, we plot $w(V_\mathrm{g})$ calculated using Eq.\ (\ref{wdef}) with $\nu_0 =p$ and employing $\delta_-$ deduced from the plateaus in the magnetic-field sweep, Fig.\ \ref{IQHsVdep}(m) \cite{deltanote}. We found $\delta_-$ is roughly around 0.2 regardless of the value of $p$. We can see that $w$ becomes larger for higher fillings, mainly due to the enhancement of $\eta = 1-\delta_- / p$ in Eq.\ (\ref{wdef}). This, in turn, enhances $C$ (see the inset of Fig.\ \ref{wnus}), resulting in the sublinear relation shown in Fig.\ \ref{IQHsVdep}(n).  Linear relation holds for $\nu_0 \leq 4$, where $w$ remains virtually unchanged. The calculated $f_0(V_\mathrm{g})$ using $a(V_\mathrm{g})$ in Fig.\ \ref{nawVgdep}(b), $w(V_\mathrm{g})$ in Fig.\ \ref{wnus}, and Eqs.\ (\ref{f0CapMod}) and (\ref{Capacitance}) are plotted with yellow lines in Figs.\ \ref{IQHsVdep}(b)--\ref{IQHsVdep}(l), which basically reproduce the experimental observed $f_0(V_\mathrm{g})$ \cite{errorf0}.

We can also calculate the variation of $f_0$ within a quantum Hall plateau using Eqs.\ (\ref{f0CapMod}), (\ref{Capacitance}), and (\ref{wdef}) by fixing $p$ to an integer value and letting $\nu_0$ vary with $B$ within the span of the plateau. The result of the calculation is plotted by thick yellow lines in Fig.\ \ref{Vm60nu34Bdep} for the quantum Hall plateaus encompassing $\nu_0 =3$ and 4, which roughly reproduce the experimentally observed behavior. With the decrease of the magnetic field $B$, the bulk filling $\nu_0$ increases, which, from Eq.\ (\ref{wdef}), leads to the decrease in $w$, and hence to the increase in $f_0$ via decreasing $C$.  

In our model, we defined the total width of the edge region containing both compressible and incompressible strips as the width of the edge state. The success in reproducing the behavior of $f_0(V_\mathrm{g})$ by our model calculation suggests that EMP is actually excited in this region straddling the two types of stripes. Kumada \textit{et al}.\ measured the velocity of EMP employing the time-of-flight method for a 2DEG covered by a metallic gate and also for an ungated 2DEG \cite{Kumada11}. They found from the analysis of the result that, while EMP straddles the incompressible strips for the ungated 2DEG, the width of EMP is truncated by the innermost (the widest) incompressible strip for the gated 2DEG\@. This is ascribed to the screening by the metallic gate of the electrostatic interaction across the incompressible strip. In our case, although a metallic gate (CE) is present relatively close to the edge state, it is not facing the 2DEG\@. The screening is therefore not strong enough to truncate the EMP by any of the incompressible strips.

\subsection{Qualitative interpretation of $f_0$ for $V_\mathrm{dpl} < V_\mathrm{g} < 0$ \label{intf0}}
\begin{figure}[t]
\includegraphics[width=8.5cm]{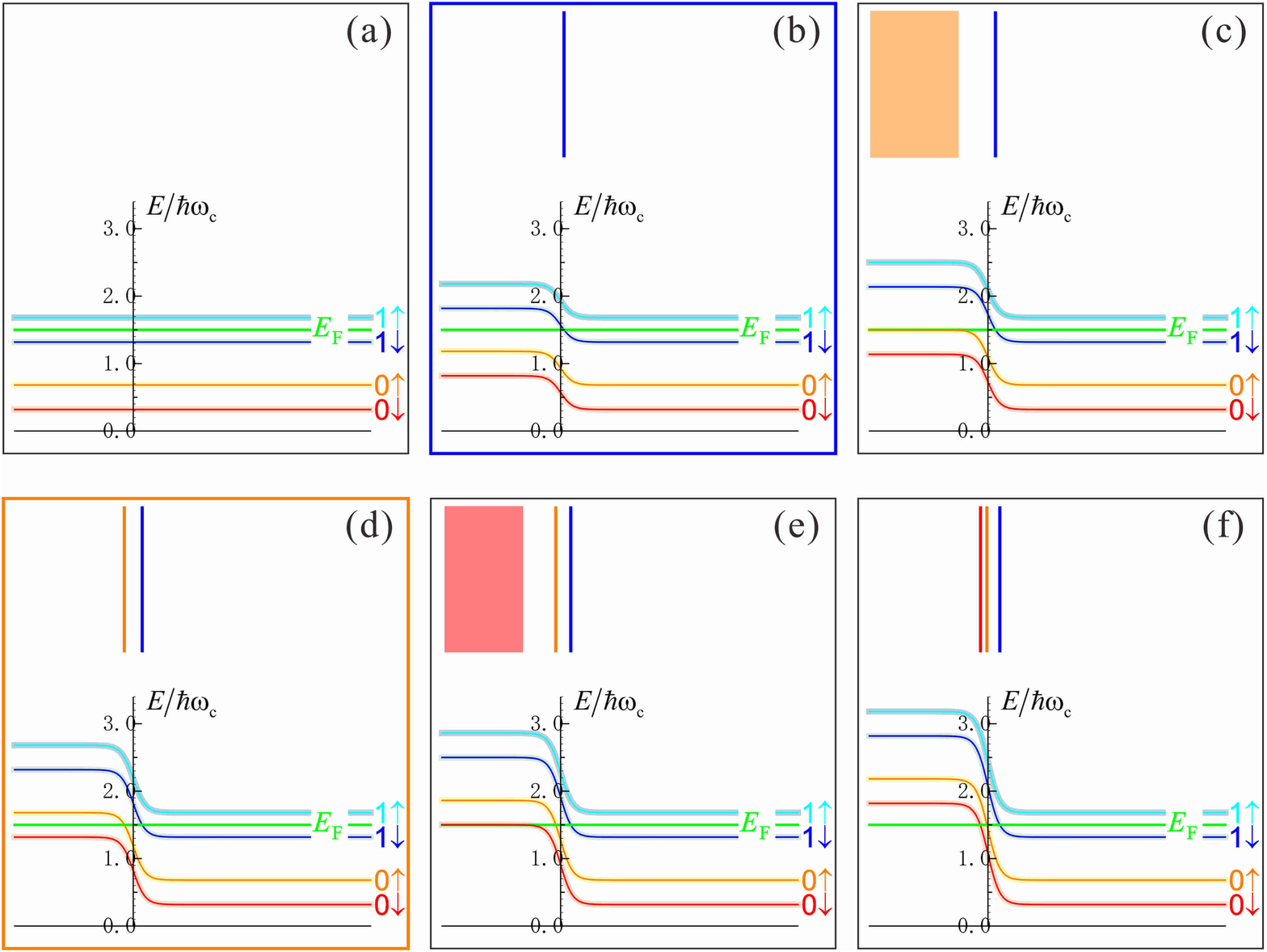}%
\caption{\label{LLE} (Color online) Landau level alignment of gated (left-hand side) and ungated (right-hand side) regions (bottom) and top view of edge and extended states at the Fermi energy $E_\mathrm{F}$ (top) for $V_\mathrm{g} = 0$ (a) and for successively more negative values of $V_\mathrm{g}$ (b) -- (f). 2DEG under the gate is fully depleted in (f). }
\end{figure}

In this subsection, we describe a qualitative explanation for the behavior of $f_0$ in the regime of $V_\mathrm{g}$ where the 2DEG beneath the CE is still not completely depleted. Starting from $V_\mathrm{g} = 0$ (Fig.\ \ref{LLE}(a)),  Landau levels initially located below the Fermi energy $E_\mathrm{F}$ shift upward and cross $E_\mathrm{F}$ one by one by applying successively more negative gate voltages (Figs.\ \ref{LLE}(b)--\ref{LLE}(f)). When $E_\mathrm{F}$ align with one of the Landau levels, or more precisely, with the areas closer to the center than the mobility edge of a disorder broadened Landau level, the region below the CE is in the extended state with mobile electrons (Figs.\ \ref{LLE}(c) and \ref{LLE}(e)). These electrons, residing on the same plane as the edge state in the slot region, add to the screening, thereby reducing $f_0$. When, on the other hand, $E_\mathrm{F}$ align with the tail of a Landau level farther away from the center than the mobility edge, and the region below the CE is in an IQH state (localized state) without any mobile electrons and thus without the additional screening (Figs.\ \ref{LLE}(b) and \ref{LLE}(d)), $f_0$ will be enhanced compared to the previous cases. The positions of the small humps observed in the plots of $\Delta \mathcal{T}(f; V_\mathrm{g})$ or $V_1(f; V_\mathrm{g})$ can thus be interpreted as the values of $V_\mathrm{g}$ at which the gated region becomes an IQH state \cite{lowDOS}. For instance, humps at around $V_\mathrm{g} = -0.20$ V and $-0.32$ V (indicated by blue and orange downward arrows, respectively, in Figs. \ref{IQHsVdep}(a) and \ref{IQHsVdep}(b)) correspond to IQH states with the fillings $\nu_\mathrm{CE} = n_\mathrm{CE} h / (eB) = 2$ and 1, respectively, of the gated region (schematically depicted in Figs.\ \ref{LLE}(b) and \ref{LLE}(d), respectively). 
In general, humps corresponding to integer values of $\nu_\mathrm{CE} < \nu_0$ are observed, so long as the IQH effect $\nu_\mathrm{CE}$ is well resolved. In addition to those mentioned above, we can observe humps attributable to $(\nu_0, \nu_\mathrm{CE}) = (2,1)$, (4,2), (4,1), (5,2) in Figs.\ \ref{IQHsVdep}(l),  \ref{IQHsVdep}(j), and \ref{IQHsVdep}(i) \cite{slant}. Vertical brighter lines, increasing in number for higher $\nu_0$, are seen in the upper panels of Fig.\ \ref{IQHsVdep}, which are presumably the remnants of the unresolved humps and thus correspond to (spin-unresolved even) IQH state in the gated region.

\section{Conclusions}
To summarize, we have shown that EMP excitations in the quantum Hall edges introduced into the slot regions of CPW can be detected as enhancement in the microwave transmission and also as the concomitant changes in the thermoelectric voltages. A series of peaks corresponding to the fundamental mode with the frequency $f_0$ and its higher harmonics at $i f_0$ ($i = 2$, 3, 4,...) are observed. A negative bias $V_\mathrm{g}$ applied to the metallic gate (CE of the CPW) to introduce the edges by depleting the underlying 2DEG also alters $f_0$. By applying a negative $V_\mathrm{g}$ beyond the depletion of the 2DEG, $f_0$ increases with the decrease of $V_\mathrm{g}$. This is mainly attributable to the effect of a more negative $V_\mathrm{g}$ to repel the edge state away from the gate, thereby reducing the capacitive coupling between them.  Among different IQH states, $f_0$ exhibits sublinear increase with the integer filling factor $\nu_0 = p$ ($p = 2$, 3, ... 18), resulting from the combined effect of $f_0 \propto \sigma_{xy} \propto p$ and the increase of the edge width $w$ with increasing $p$. Within an IQH plateau, $f_0$ increases with decreasing magnetic field owing to the decrease of $w$. The experimentally observed behaviors of $f_0$ can be reproduced well by the calculations based on the computed profile $n(x)$ of the electron density in the vicinity of the edge. This confirms the accuracy of the theory used in the computation\cite{Larkin95} and, at the same time, demonstrates that the measurement of $f_0$ can be a useful tool to explore the profile of the edges. 

We have also found that EMP takes place even when the 2DEG under the gate is not completely depleted, with a lower $f_0$ compared to the fully depleted cases. Slight shift of $f_0$ to higher-frequency side is observed when the 2DEG below the gate is in the localized IQH state with reduced screening.

\begin{acknowledgment}
This work was supported by JSPS KAKENHI Grant Numbers JP26400311 and JP17K05491.
\end{acknowledgment}

\appendix
\section{Derivation of Eq. (\ref{f0CapMod}) \label{Af0}}
In this appendix, we derive Eq.\ (\ref{f0CapMod}), basically combining the procedures in Refs.\ \citen{Volkov88} and \citen{Zhitenev95}. We define the 2DEG plane as the $x$-$y$ plane with $x$-axis ($y$-axis) across (along) the edge (see Fig.\ \ref{nxdep}). Assuming translational invariance along the $y$-axis, the continuity equation results in
\begin{equation}
e\frac{\partial n(x)}{\partial t} = \frac{\partial j_x}{\partial x} \approx \frac{\partial}{\partial x} (\sigma_{xy} E_y), \label{continuity}
\end{equation}
where we used the relation $\sigma_{xx} \ll \sigma_{xy}$ between the diagonal and the Hall conductivity in approximating the current density in the $x$-direction, $j_x$, and $E_y = -\partial \phi / \partial y$ is the electric field in the $y$-direction with $\phi = \phi_0 \exp(i k_y y -i \omega t)$ the EMP potential. In the capacitance approximation, we assume that the linear charge density $Q = Q_0 \exp(i k_y y -i \omega t)$ is related to $\phi$ by the capacitive coupling
\begin{equation}
Q = C \phi, \label{phiC}
\end{equation}
where $C$ is the capacitance per unit length. Integrating Eq.\ (\ref{continuity}) across the width of the edge region, we have 
\begin{equation}
\int_a^{a+w} e\frac{\partial n(x)}{\partial t} dx = \frac{\partial Q}{\partial t} = -\sigma_{xy} \frac{\partial \phi}{\partial y}. \label{wave}
\end{equation}
From Eqs.\ (\ref{wave}) and (\ref{phiC}), we obtain 
\begin{equation}
\omega = k_y \frac{\sigma_{xy}}{C},
\end{equation}
which is equivalent to Eq.\ (\ref{f0CapMod}) noting that $\omega = 2 \pi f_0$ and $k_y = 2 \pi /L$ \cite{Wavelength}.

\bibliography{ourpps,lsls,twodeg,emp,qhe,misc,wc,ninehlvs,thermo,noteEMPQHS}

\begin{thebibliography}{10}
\bibitem{2DEGdepth}
Here, we defined the depth $d$ of 2DEG as the distance from the surface to the
  maximum of the envelope of the 2DEG wavefunction, which is the addition of
  the depth of the heterointerface, 60 nm, and the distance of the maximum from
  the interface, 5nm. \cite{Endo05MA}.
\bibitem{contactpot}
We estimate that the metallic gates in our sample have a contact potential
  $-$0.025 V lower than that of the 2DEG, which is taken into consideration in
  the calculation of $V_\mathrm{g}$ dependence of $f_0$, $n(x)$, $a$, and $w$.
\bibitem{pinned}
We have also repeated the following calculations using ``pinned'' surface
  model. This resulted in much worse agreement between the calculated and the
  experimentally obtained $f_0(V_\mathrm{g}$), attesting to the appropriateness
  of the ``frozen'' surface model in describing our experiments.
\bibitem{deltanote}
Strictly speaking, $\delta_-$ determined by varying $B$ for a fixed electron
  density $n_0$ and that for a fixed $B$ and varying $n(x)$ can be slightly
  different. In the present paper, however, we neglect the difference and
  approximate $\delta_-$ by half the width of the plateau observed in the
  magnetic-field sweeps (Figs. \ref{IQHsVdep}(m) and \ref{Vm60nu34Bdep}).
\bibitem{redisbulk}
In Eq.\ (\ref{wdef}), we used $n(x)$ at $B = 0$ T and neglected the possible
  effect of the magnetic-field induced redistribution of the electrons around
  the border between the edge and the bulk regions on the width $w$ of the edge
  state. Furthermore, the total width of the incompressible stripes should be,
  strictly speaking, subtracted from $w$, which we neglected noting that the
  incompressible stripes are much narrower than the compressible stripes.
  \cite{Chklovskii92}.
\bibitem{errorf0}
The calculated traces of $f_0(V_\mathrm{g})$ for higher fillings exhibit slight
  downturn when decreasing $V_\mathrm{g}$ close to $-0.6$ V, which is absent in
  the experimentally observed $f_0(V_\mathrm{g})$. We presume that the downturn
  results from overestimate of $w$ in Fig.\ \ref{wnus}. For higher fillings,
  $\eta$ in Eq.\ (\ref{wdef}) closely approaches unity, and a slight error in
  any parameter ends up in considerable error in $w$ determined by the
  equation.
\bibitem{lowDOS}
Note that owing to the small density of states at the tails of Landau
  levels,$V_\mathrm{g}$ dwells at the IQH (localized) states for only short
  span in the $V_\mathrm{g}$ sweep.
\bibitem{slant}
We can recognize slightly slanted bright stripes running nearly parallel to the
  humps in the plots of $\Delta \mathcal{T}(f,V_\mathrm{g})$ and
  $V_1(f,V_\mathrm{g})$. We currently have no explanation for these stripes.
  Although these stripes make it difficult to perceive, the positions of the
  humps for higher harmonics remain unchanged from that of the fundamental
  frequency $f_0$.
\bibitem{Wavelength}
Note that the wavelength of the microwave is orders of magnitude longer than
  $L$. Therefore, the amplitude of the microwave is virtually uniform across
  the sample, and $k_y$ is solely determined by the length $L$ along which the
  2DES is affected by the microwave.
\end{thebibliography}


\end{document}